# DMSO-Induced Dehydration of DPPC Membranes Studied by X-ray Diffraction, Small-Angle Neutron Scattering, and Calorimetry

M. A. Kiselev*, P. Lesieur#, A. M. Kisselev*, C. Grabielle-Madelmond+, M. Ollivon+

* Frank Laboratory of Neutron Physics, JINR, 141980 Dubna, Moscow region, Russia
# LURE, Université Paris-Sud, Bât. 209-D, F91405 Orsay cedex, France
+ Physico-Chimie des systèmes polyphasés, URA 1218 du CNRS, Faculté de Pharmacie, tour B, F – 92296, Chatenay Malabry, France

**ABSTRACT**

The influence of dimethyl sulfoxide (DMSO) on membrane thickness, multilamellar repeat distance, and phase transitions of 1,2-dipalmitoyl-sn-glycero-3-phosphatidylcholine (DPPC) was investigated by X-ray diffraction and small-angle neutron scattering (SANS). The differential scanning calorimetry (DSC) study of water freezing and ice melting was performed in the ternary DPPC/DMSO/water and binary DMSO/water systems. The methods applied demonstrated the differences in membrane structure in three sub-regions of the DMSO mole fraction ($X_{DMSO}$): from 0.0 to 0.3 for the first, from 0.3 to 0.8 for the second, and from 0.9 to 1.0 for the third sub-region. The thickness of the intermembrane solvent at $T = 20°C$ decreases from $14.4 \pm 1.8$ Å at $X_{DMSO} = 0.0$ to $7.8 \pm 1.8$ Å at $X_{DMSO} = 0.1$. The data were used to determine the number of free water molecules in the intermembrane space in the presence of DMSO. The results for $0.0 \leq X_{DMSO} \leq 0.3$ were explained in the framework of DMSO-induced dehydration of the intermembrane space.

*Correspondence to* :
M. A. Kiselev - Frank Laboratory of Neutron Physics, Joint Institute for Nuclear Research, 141980 Dubna, Moscow region, Russia.
E-mail: kiselev@nf.jinr.ru, fax: 7-096-21-65882

1. INTRODUCTION

Dimethyl sulfoxide has two very important biological properties: a) ability to protect a variety of cells from the damaging effects of freezing and storage at a very low temperature, b) the modification of X-ray induced damage in cells and whole animals when DMSO is present before and during exposure to radiation [1].

The main purpose of cryobiology is to find an "optimal" way of cooling biological systems to low temperatures (about the temperature of liquid nitrogen) and, at the same time, prevent the formation of ice inside the biological tissue. The mechanism of DMSO cryoprotection is still debated and, on the molecular level, it is not clear [2].

DMSO makes hydrogen bonds with water molecules. The structure of the binary DMSO/water system has been studied by means of the spin–lattice relaxation and the chemical shift behavior of water and DMSO protons [3]. It was found that DMSO and water molecules tend to form hydrogen bonds in the relation 1/2 (mole DMSO fraction $X_{DMSO} = 0.33$) or 1/3 ($X_{DMSO} = 0.25$). The DMSO/water phase diagram is well known [4] and the phase behavior of the binary DPPC/water system has been investigated by calorimetry [5]. The influence of DMSO fraction on the pre–transition existence and on the repeat distance of the DPPC multilamellar structure has been studied for the region of a small DMSO mole fraction, $X_{DMSO} \leq 0.13$. From the calculation of the electron density profile, it was established that the intermembrane solvent space decreases with the increasing DMSO concentration [6].

In the present paper, the influence of DMSO on 1,2-dipalmitoyl-sn-glycero-3-phosphatidylcholine (DPPC) membranes was studied by X-ray diffraction, small-angle neutron scattering (SANS), and differential scanning calorimetry (DSC). Our purpose was to obtain information on changes in the structure of DPPC in the presence of DMSO at T=20°C and during the formation of ice. Calorimetric and time resolved X-ray diffraction measurements were performed in order to explore the phase transitions in the DPPC/DMSO/water system for $0.0 \leq X_{DMSO} \leq 1.0$.

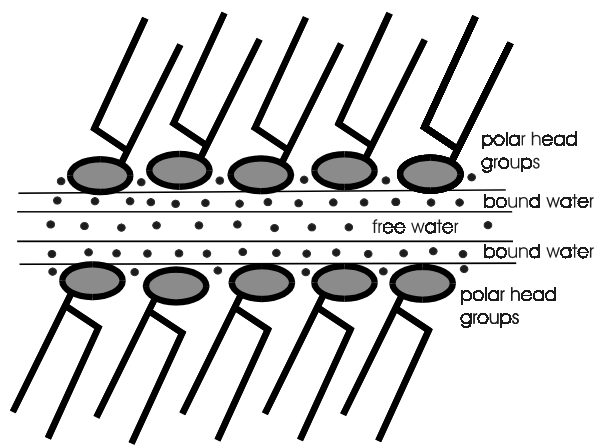

FIGURE 1. Arrangement of water molecules in the phospholipid/water system. Water molecules are divided into three fractions: the first – free water molecules located in the center of the intermembrane space; the second – bound water molecules located in the intermembrane space near the polar heads of DPPC molecules; the third - bound water molecules located in the spatial region of DPPC head groups.

The phospholipid/water multilayer system (multilamellar membrane) can be represented as described in Figs. 1 and 2. Fig. 1 presents the model for the water arrangement in the intermembrane space. It is possible to divide water molecules into two fractions: a) bound water – where water molecules are bonded to the polar head groups of the phospholipid molecules; b) free water molecules, which exist between lipid bilayers without any interaction with membrane surface. The bound water can be further divided into the fraction of molecules embedded in the spatial region of the polar head groups and the fraction located in the intermembrane space near the polar head groups. The number of bound water molecules located in the region of the polar head groups and in the intermembrane space is equal to 2.2 and 8.6 per DPPC molecule, respectively; the number of free water molecules per DPPC molecule is equal to four [5, 7].

## 2. MATERIALS

1,2-dipalmitoyl-sn-glycero-3-phosphatidylcholine (DPPC, > 99% purity) was purchased from Sigma corporation (Paris, France). Spectrophotometric grade dimethyl sulfoxide (DMSO, > 99% purity) and deuterated DMSO (99.9%) were purchased from Aldrich (Paris, France). Water was of Millipore standard (18 MΩ · cm).

## 3. SAMPLE PREPARATION

The samples were prepared either in eppendorfs for calorimetry or in capillaries for X-ray diffraction. The X-ray quartz capillaries (1.5 mm in diameter with 0.01 mm wall thickness) were purchased from GLAS company (W.Muller, Berlin, Germany). The desired amount of lipid and DMSO / water solution was placed in the containers. Both, eppendorfs and capillaries were hermetically sealed with paraffin. The samples were heated to a temperature above the main phase transition temperature and sonicated in the Branson sonicator for 5 min. After sonication, the samples were stored at a temperature above the main phase transition temperature for about 3 hours and then cooled to room temperature. The heating/cooling process was repeated 4 times in order to obtain a homogeneous DPPC multilamellar suspension, which is necessary for calorimetry and for powder diffraction studies of the membrane structure [8]. One drop of the sample was placed in each sealed aluminium platelet for the differential scanning calorimetry measurements. The sample weight in the platelets was 10 to 15 mg.

The weight ratio between DPPC and solvent was chosen to keep the number of solvent molecules per DPPC molecule constant. Let us denote the number of DMSO, water, and DPPC molecules as $N_{DMSO}$, $N_w$, and $N_{lip}$, respectively. We denote the weights of the solvent and DPPC in the sample as $W_{sol}$ and $W_{DPPC}$. For DPPC in pure water, the ratio $N_w / N_{lip} = 40.8$ corresponds to the ratio $W_{sol} / W_{DPPC} = 1$. For DPPC in the DMSO / water mixture, the ratio $(N_{DMSO} + N_w) / N_{lip}$ was kept equal to 40.8 for all samples used in the diffraction and calorimetry study. This means that the solvent concentration in the sample $W_{sol} / (W_{sol} + W_{DPPC})$ is larger than 0.5 for all DMSO concentrations. All measurements were performed with an excess of the solvent. The mole fraction of DMSO $X_{DMSO}$ is defined as the mole fraction of DMSO in the solvent mixture $N_{DMSO} / (N_{DMSO} + N_w)$.

For the SANS study, we used large unilamellar vesicles prepared by the extrusion of heated (50-60°C) 1% (w/w) DPPC dispersion through Nuclepore membrane filters (pores 2000 Å in diameter) [9].

## 4. EXPERIMENTAL TECHNIQUES

The X-ray diffraction experiment was carried out at the D-22 and D-24 spectrometers of the DCI synchrotron ring. Fig. 3 presents the X-ray diffraction pattern for the DPPC/water system at $T = 20°C$ measured at the D-24 spectrometer. The sample-to-detector distance was chosen to be small enough (314mm) to measure the small-angle diffraction pattern from the multilamellar DPPC structure (the region of scattering vector $q$ of $0.0 \text{ Å}^{-1} \leq q \leq 0.5 \text{ Å}^{-1}$) and to also measure, at the same time, the wide-angle diffraction peaks from the hydrocarbon chains of DPPC ($q \approx 1.5 \text{ Å}^{-1}$). The five diffraction peaks from the multilamellar DPPC structure correspond to the membrane repeat distance $d = 63.7 \pm 3.2 \text{ Å}$ (see Fig. 2).

In the region of $q \approx 1.5$ Å$^{-1}$, the diffraction pattern is interpreted as diffraction on a quasihexagonal lattice from the hydrocarborn chains. The sharp reflex at $q = 1.50$Å$^{-1}$ corresponds to the lattice constant $4.19 \pm 0.06$Å, the "shoulder" on the side of large $q$ is an additional wide diffraction peak ($q = 1.53$Å$^{-1}$) from the lattice constant $4.11 \pm 0.06$Å. The diffraction pattern on Fig. 3 contains a full information about lamellar and lateral DPPC membrane structures, nevertheless this geometry results in a large uncertainty in the determination of the repeat distance (about 5 %) compared with the uncertainty in the determination of the lattice constants for hydrocarbon chains (about 1.5 %). An increase in sample-to-detector distance reduces the experimental errors in the determination of $d$, which is necessary for the phase transitions investigation in the DPPC/DMSO/water system. For this reason, the sample-to-detector distance $L_{S-d}$ was chosen to be 788mm, which gives an uncertainty in the determination of the repeat distance of 1.2 % - 1.4 % for different DPPC phases.

The D22 spectrometer was chosen because it had better resolution than the D24 spectrometer. With $L_{S-d} = 788$mm. and with direct beam location in the center of the linear position sensitive detector, only the first order diffraction was detected from the DPPC multilamellar structure. Fig. 4 shows the X-ray diffraction patterns measured at the D22 spectrometer. The repeat distance $d$ of the multilamellar structure was determined from the position of the diffraction peak using the Bragg equation $2d \sin\theta = n\lambda$, where $\theta$ is the half of the scattering angle and $n$ is the order of the diffraction peak, which is $n=1$ in our experiment.

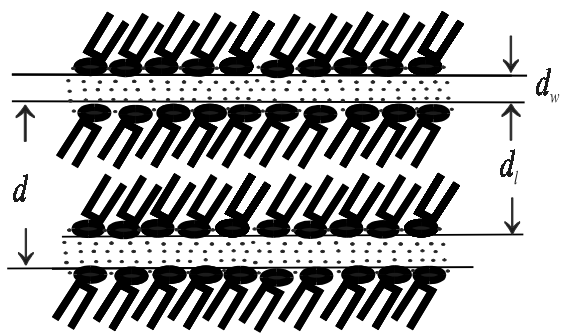

FIGURE 2. Schematic representation of the gel $L_{\beta'}$ - phase of DPPC multilamellar structure. The multilamellar structure consist of the DPPC bilayers with membrane thickness $d_l$ and solvent layers with solvent thickness $d_w$. The repeat distance of DPPC membrane $d = d_l + d_w$. In the $L_{\beta'}$ - phase, the DPPC hydrocarbon chains are tilted to the membrane surface.

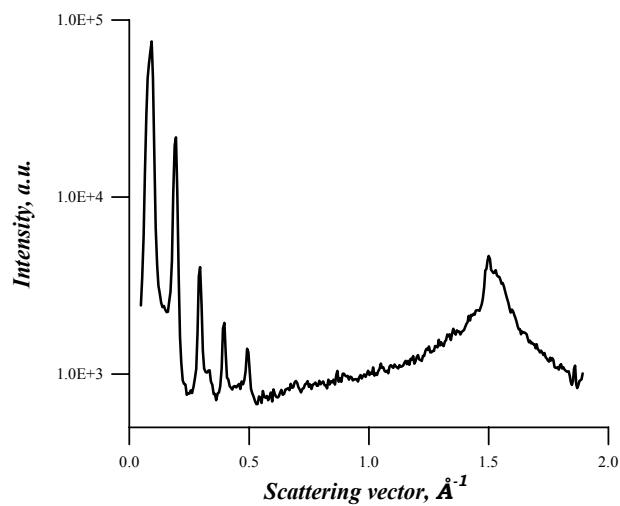

FIGURE 3. The diffraction pattern from DPPC/water system at $T = 20°C$. Five diffraction peaks in the region of scattering vector $0.0$ Å$^{-1} \leq q \leq 0.5$ Å$^{-1}$ correspond to the powder diffraction with n=1,2,3,4,5 from the multilamellar structure of DPPC with $d = 63.7 \pm 3.2$Å. Two overlapped peaks in the region of $q \approx 1.5$ Å$^{-1}$ correspond to diffraction from the mutual packing of the hydrocarbon chains within the bilayer.

SANS measurements were performed at the YuMO time-of-flight small-angle spectrometer (pulse neutron source IBR-2) in the Frank Laboratory of Neutron Physics. Guinier approximation of the scattering cross-section was used to interpret the neutron small-angle scattering spectra [10, 11]. For unilamellar vesicles with a surface area $S$, thickness $d_l$, and radius of gyration $R_t$, the approximation that is valid for scattering vectors $q$ in the domain $2\pi/\sqrt{S} < q < 1/R_t$, is given by equation

$$\frac{d\sigma}{d\Omega} = \frac{2\pi S (\Delta\rho\, d_l)^2}{q^2} \exp[-q^2 R_t^2] \ , \qquad (1)$$

where $\Delta\rho$ is the scattering length density relative to the solvent. For a homogeneous membrane, the membrane thickness $d_l$ is calculated by the formula $d_l = R_t \sqrt{12}$ [10, 11, 12].

Differential thermal analysis measurements were performed using the Perkin Elmer DSC 4 calorimeter. For investigations of the main phase transition and pre-transition, the temperature interval from 0°C to 85°C and the scanning rate of 2.0 degrees per minute were used. The phase transition of lauric acid ($T_{ph}$ = 43.7°C, $\Delta H$ = 42.58 cal / g) was used for calibration. To study the phase transitions associated with the freezing of water and melting of ice, the temperature interval from 20°C to –130°C and the scanning rate of 10 degrees per minute were used. In this case, we used the melting temperature of pure water for calibration.

## 5. RESULTS AND DISCUSSION
### 5.1. Influence of DMSO on the Structure and Main Phase Transitions

Fig. 4 depicts the sequence of diffraction patterns recorded with an acquisition time of 2 min on heating the sample from $T = 25$°C to $T = 80$°C. The heating rate 1°C/min was used in the temperature ranges 25°C - 40°C and 50°C `- 80°C. In the temperature range from 40°C to 50°C, the heating rate was decreased down to 0.5°C/min, which gave us a 1°C step between the diffraction patterns. For precise measurements, the heating rate of 0.2°C/min was used in the region near the main phase transition in order to obtain a more accurate determination of the phase transition temperature. This gave us an accuracy of ± 0.4°C for the determination of the phase transition temperature.

In the temperature interval 25°C - 33°C, the diffraction patterns (see Fig. 4) correspond to the gel $L_{\beta'}$ - phase. From $T = 35$°C to $T = 41$°C, the membrane exhibits the ripple phase $P_{\beta'}$, which is characterized by a shift of the diffraction peak to the region of smaller $q$, by a decrease in the diffraction peak intensity, and by peak broadening. The origin and structure of $P_{\beta'}$ - phase is still debated. Nevertheless, the membrane undulations are a common $P_{\beta'}$ - phase property [13]. At $T = 41.5$°C, the membrane undergoes a phase transition from the $P_{\beta'}$ - phase to the liquid crystalline $L_\alpha$ - phase, the so-called main phase transition [14]. In the temperature interval of 42°C $\leq T \leq$ 80°C the diffraction pattern correspond to the liquid crystalline $L_\alpha$ - phase, which is characterized by a melted hydrocarbon chains.

As the DMSO mole fraction increases, the region of $P_{\beta'}$ - phase decreases and membrane undulations disappear at $X_{DMSO}$ = 0.10. At the same time, the temperature of the main phase transition increases with increasing DMSO mole fraction. The same result was obtained by differential thermal analysis. Fig. 5 shows the DSC curves for six samples with different DMSO mole fractions from $X_{DMSO} = 0.00$ to $X_{DMSO} = 0.10$. For example, for the curve recorded from the DPPC /water system, the first endothermic peak (so called pre-transition) at a temperature of about 37°C corresponds to the transition from the $L_{\beta'}$ - phase to the $P_{\beta'}$ - phase. The second endothermic peak corresponds to the main phase transition from the $P_{\beta'}$ - phase to $L_{\alpha}$ - phase. The difference between the pre-transition and the main phase transition temperatures diminishes while $X_{DMSO}$ increases. At $X_{DMSO} = 0.10$, two endothermic peaks merge together and the pre-transition peak disappears. The same phenomenon has been recently observed [6], but with a different concentration ($X_{DMSO} = 0.04$) for the pre-transition disappearance.

Fig. 6 presents the dependence of the main phase transition temperature $T_{ch}$ and of the repeat distance $d$ on the mole DMSO fraction at $T = 20$°C measured by X-ray diffraction and calorimetry. The values of $T_{ch}$ obtained by X-ray diffraction and by calorimetry are in good agreement. We can divide the range of the DMSO fractions into three regions. The first is that of small DMSO mole fractions ($0.0 \leq X_{DMSO} \leq 0.3$), where important changes in the $d$ and the $T_{ch}$ are observed. The second is the region of intermediate DMSO mole fractions ($0.3 \leq X_{DMSO} \leq 0.8$), where $d$ and $T_{ch}$ are constant. In the third region, near pure DMSO solvent, the DPPC membrane undergoes a transition towards a phase characterized by a small repeat distance of 52 Å and a high transition temperature of $74.2 \pm 0.2$°C.

To understand the described results for the ternary DPPC/ DMSO/ water system, it is important to note that the absence of the pre-transition and increased phase transition temperature are properties of the partly dehydrated DPPC membranes in the case of the binary DPPC/ water system [5].

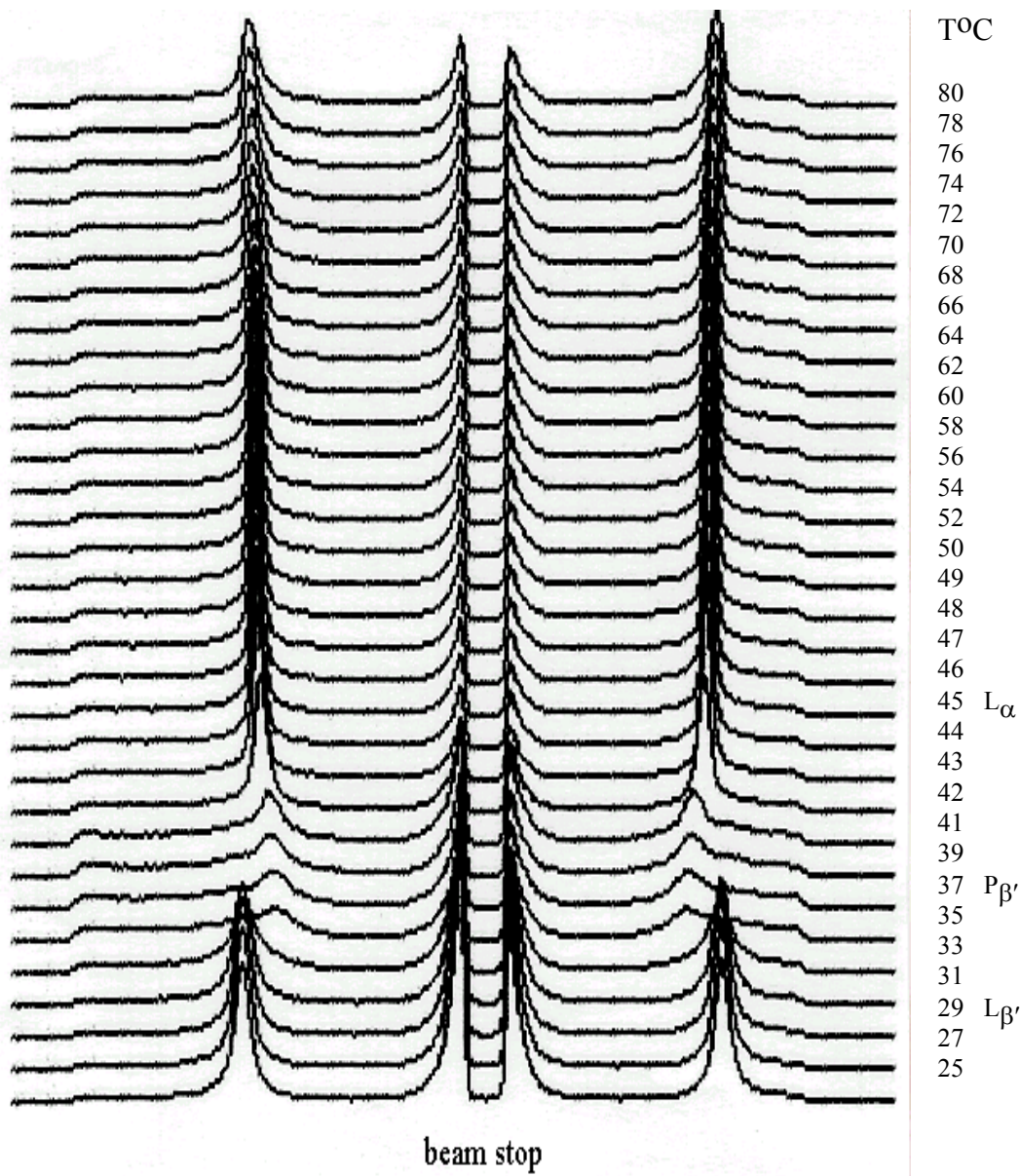

FIGURE 4. The sequence of diffraction patterns from the DPPC/water system recorded with an acquisition time of 2 min when heating the sample from $T = 25°C$ to $T = 80°C$. The intensity oscillations in the center of the figure correspond to the direct beam absorption by the beam stop. The left and right peaks symmetrical to the center of the beam correspond to the first-order powder diffraction from the multilamellar structure. Gel $L_{\beta'}$ - phase exist in the temperature interval of $25°C \leq T \leq 33°C$, the ripple phase $P_{\beta'}$ in the interval of $35°C \leq T \leq 41°C$, and liquid crystalline $L_\alpha$ - phase in the interval of $42°C \leq T \leq 80°C$.

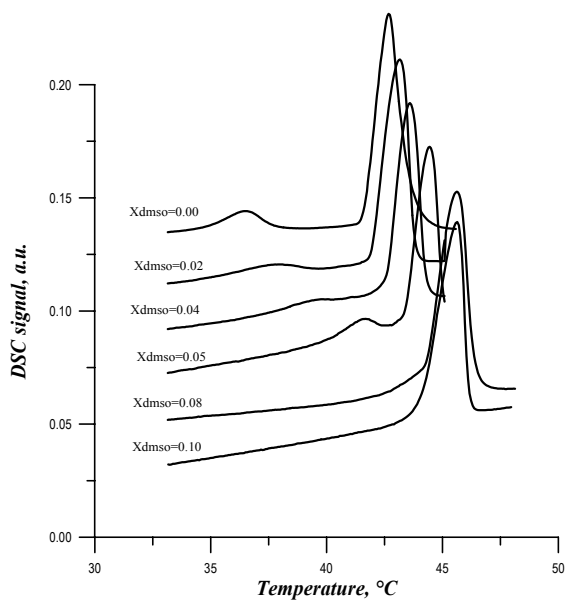

FIGURE 5. Calorimetric curves for the main phase transition in the DPPC /DMSO /water system (long sharp peaks) at $X_{DMSO} = 0.0$, 0.02, 0.04, 0.05, 0.08, and 0.1, and pre-transition (broad peaks on the left side of the curves for $X_{DMSO} = 0.0$, 0.02, 0.04, and 0.05). The ordinate axis is in relative unites. The pre-transition peaks disappear after $X_{DMSO} = 0.08$.

The repeat distance $d$ of the DPPC multilayer is connected with the membrane thickness $d_l$ and thickness of the solvent $d_w$ in the intermembrane space via the expression $d = d_l + d_w$. A decrease in the DPPC membrane repeat distance at $T = 20°C$ can occur as a result of $d_l$ or $d_w$ decrease. The SANS experiment with large unilamellar DPPC vesicles was carried out to determine the influence of DMSO concentration on the bilayer thickness $d_l$. The values of $d_l$ obtained at $T = 20°C$ with an accuracy of $\pm 1$ Å in the Guinier approximation (1) are 49.6 Å for $X_{DMSO} = 0.0$, 49.2 Å for $X_{DMSO} = 0.05$, and 50.1 Å for $X_{DMSO} = 0.10$. The same results concerning the absence of DMSO influence on the membrane thickness were obtained recently from the electron density profile of the DPPC /DMSO /water system for $0.0 \leq X_{DMSO} \leq 0.13$ [6]. This important fact obtained by two different techniques applies to the interaction of DMSO with respect to the region of the polar head groups. The DMSO molecules do not penetrate to the region of the polar head groups of DPPC membrane and, thus, have no influence on the membrane thickness.

Fig. 7 demonstrates SANS curves for DPPC vesicles in the deuterated DMSO /$D_2O$ solvent. The scattering curves from vesicles in pure $D_2O$ and from vesicles in deuterated DMSO /$D_2O$ solvent with $X_{DMSO} = 0.1$ practically coincide. As can be seen from Fig.6, the region $0.0 \leq X_{DMSO} \leq 0.10$ is the region where the main changes in the DPPC membrane repeat distance occur. The $d$ decreases from $64.0 \pm 0.8$ Å at $X_{DMSO} = 0.0$ to $57.9 \pm 0.8$ Å at $X_{DMSO} = 0.1$ and, with further increases in $X_{DMSO}$, the repeat distance diminishes slowly to a constant value of about 57 Å. These changes in the membrane structure correspond to a decrease in the intermembrane solvent thickness $d_w$ from $14.4 \pm 1.8$ Å at $X_{DMSO} = 0.0$ to $7.8 \pm 1.8$ Å at $X_{DMSO} = 0.1$, which can be explained as a reduction of the solvent amount in the intermembrane space.

The values of $d = 64.0 \pm 0.8$ Å and $d_l = 49.6 \pm 1.0$ Å obtained in our work for the DPPC /water system at $T = 20°C$ are in good agreement with the values measured by X-ray diffraction ($d$) and calculated from the electron density profile of the DPPC membrane ($d_l$): $d = 63.7 \pm 0.3$ Å and $d_l = 49.8 \pm 2.3$ Å [15].

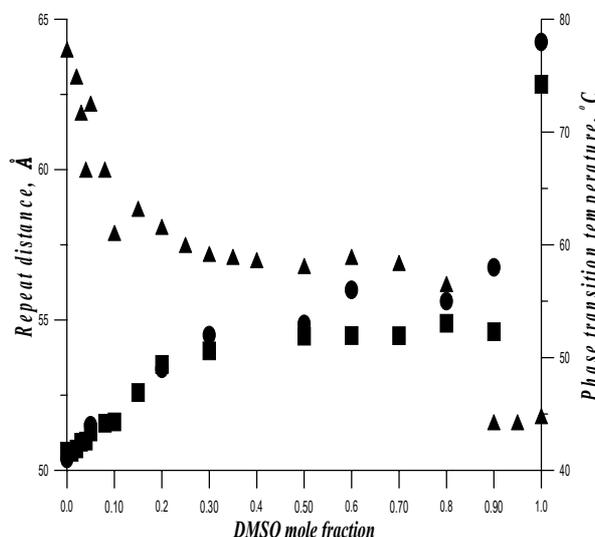

FIGURE 6. The dependence of the main transition temperature and the repeat distance of the DPPC membrane on the mole fraction of DMSO in the excess solvent. ● - main phase transition temperature obtained from the diffraction measurements with the heating rate of 0.2°C/min and acquisition time of 2 min per spectrum, ■ - main phase transition temperature obtained from the calorimetric measurements with the heating rate of 2.0°C/min, ▲ - repeat distance of multilamellar liposomes at $T = 20$°C.

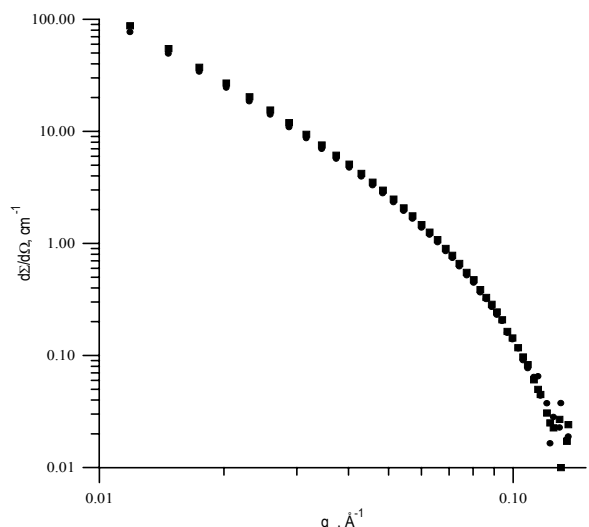

FIGURE 7. SANS curves for the extruded DPPC vesicles at $T = 20$°C: ● - in $D_2O$ solvent, $R_t = 14.3 \pm 0.3$ Å; ■ - in deuterated DMSO/$D_2O$ solvent with $X_{DMSO} = 0.1$, $R_t = 14.5 \pm 0.3$ Å. The macroscopic cross-section $d\Sigma/d\Omega$ proportional to $q^{-2}$ at small values of the scattering vector $q$. The $q^{-2}$ law corresponds to scattering from large unilamellar vesicles.

## 5.2. Crystallization of water and melting of ice

Fig. 8a presents the calorimetric curves ($X_{DMSO} = 0.0, 0.01, 0.03, 0.05, 0.08, 0.1,$ and 0.15) for the binary DMSO/water system on cooling from 20°C to −130°C with the rate of −10°C / min. For the binary DMSO/water system, only one exothermic transition is observed in this temperature range. The shape of the transition remains narrow up to $X_{DMSO} = 0.1$, but for larger cryoprotector concentrations, $X_{DMSO} = 0.15$, the transition becomes broader, as can be noted from Fig. 8a. The multiplication factor $k$ was introduced to scale the curves such that the amplitude of the transition remains approximately constant. This factor was found to be necessary because of the fast decrease in the transition enthalpy when DMSO is added. The transitions for solutions with $X_{DMSO} \leq 0.10$ correspond to the creation of a binary system consisting of an ice and liquid DMSO/water mixture. The eutectic temperature for the DMSO/water system is $-63° \pm 1$°C [4]. Consequently, the exothermic transition in the solution with $X_{DMSO} = 0.15$ at $T = -74.5$°C corresponds to the formation of ice and solid DMSO·3 $H_2O$ hydrate.

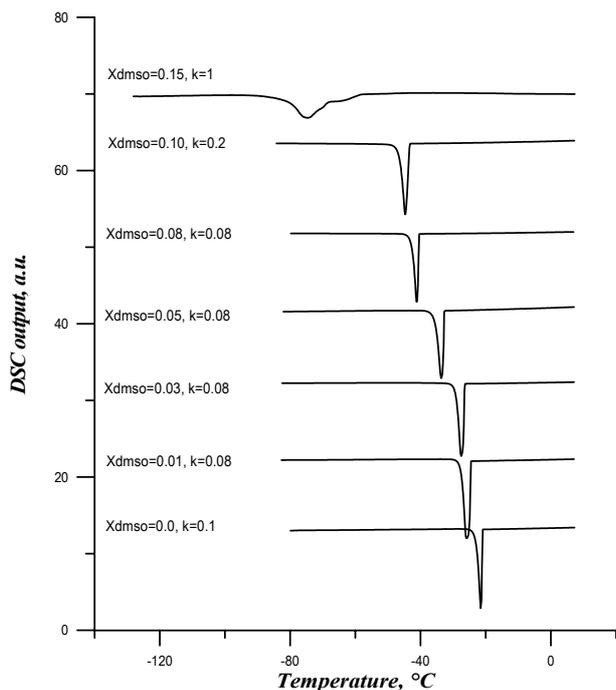

FIGURE 8a. DSC recordings of the curves upon cooling the DMSO /water mixtures from 20°C to –120°C with the rate of –10°C/min. All peaks are normalized to the peak height (normalized coefficients $k$ are presented). The ordinate axis is in the relative unites. The exothermic phase transition at $X_{DMSO} \leq 0.1$ corresponds to water crystallization. The cooling event at $X_{DMSO} = 0.15$ corresponds to water crystallization and eutectic trihydrate DMSO. Here, the output DSC signal becomes smaller and the peak broader.

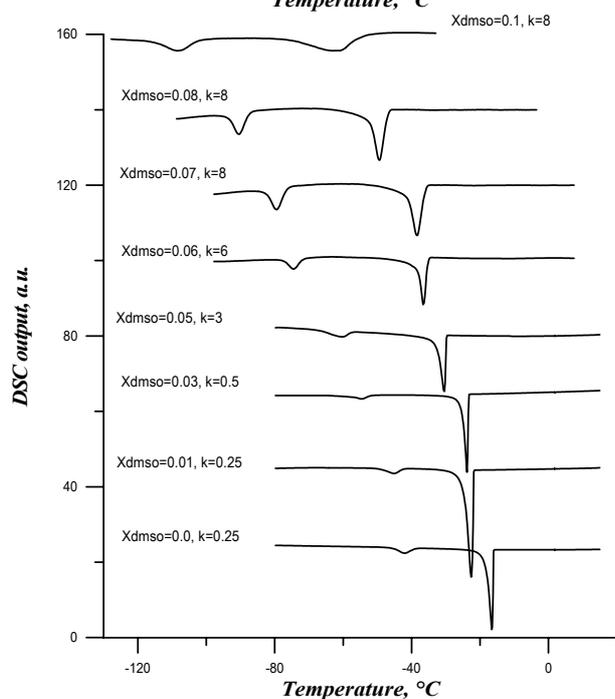

FIGURE 8b. DSC recordings of the curves upon cooling DPPC /DMSO /water mixtures under the same conditions as in Fig. 8a. All peaks are normalized to the peak height (normalized coefficients $k$ are presented). The ordinate axes is in the relative unites. Two exothermic phase transitions were measured for all DMSO concentrations. At $X_{DMSO} = 0.1$, both phase transitions become similar and difficult to detect because of a weak output DSC signal. The first transition corresponds to the formation of ice in the bulk solvent, the second to the homogeneous nucleation in the intermembrane space.

Fig. 8b shows the calorimetric curves ($X_{DMSO}$ = 0.0, 0.01, 0.03, 0.05, 0.06, 0.07, 0.08, and 0.1) for the ternary DPPC /DMSO /water systems obtained under the same conditions as described above for the binary system. Two exothermic transitions are observed at cooling from $T = 20°C$ to $T = –130°C$. The first (the temperature is close to the transition temperature of the solvent) can be attributed to the ice formation in the DMSO /water solvent (ice formation out of the intermembrane space in the bulk solvent). The second transition (the temperature is from 20 to 45°C lower than that of the first one) can be attributed to the freezing of the solvent trapped in the bilayers [5]. A multiplication factor was also introduced here to scale the data. It can be noticed that these two transitions have distinct amplitude variations with the DMSO concentration. While the amplitude of the first transition is kept roughly constant, the relative importance of the second increases when DMSO is added.

According to our results and the phase diagram of DMSO/water system [4], the ratio between the first and the second exothermic transitions in the ternary DPPC/DMSO/water system corresponds to a decrease in the amount of ice created outside the intermembrane space compared to the amount of ice created within the intermembrane space. For $X_{DMSO} \geq 0.06$, the second phase transition corresponds to the eutectic transition.

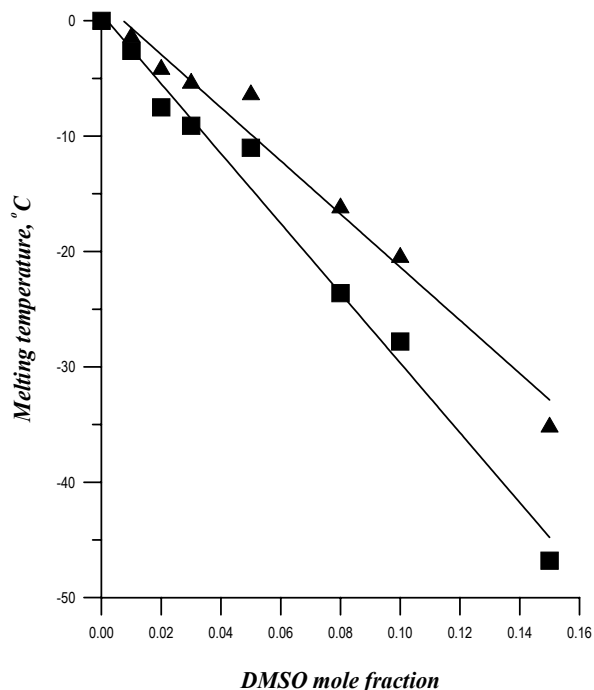

FIGURE 9. The dependence of the melting temperature of ice on the DMSO mole fraction for the DPPC/DMSO/water system, ■ pixel and for DMSO/water solvent, ▲ pixel. The melting point was determined as the offset temperature on DSC curves.

Characteristics of the studied systems that are important for future discussion were obtained from the DSC curves recorded during heating. The melting temperature of ice was determined as the offset temperature (the intersection point of the baseline with the tangent to the back side of the endothermic peak). The offset temperature was chosen because it corresponds to the end of the melting process, such that the composition of the liquid (mole DMSO fraction) in the ternary system was equal to the system composition without ice. Fig. 9 presents the differences in the dependence of the melting temperature on the DMSO mole fraction for the binary and ternary systems. For each DMSO mole fraction, the melting temperature of the DPPC/DMSO/water system is lower than the melting temperature of the DMSO/water solvent. These differences can be attributed to different DMSO mole fractions in the binary system and in the bulk solvent of the ternary system. The explanation of this phenomenon consists in the fact that some fraction of water molecules is connected with the polar head groups of the membrane and is not mixed with the DMSO molecules in the ternary DPPC/DMSO/water system.

As a result, the actual DMSO concentration $X'_{DMSO}$ in the bulk solvent of the ternary system will be higher compared with the DMSO/water solvent used for the ternary system preparation. It is possible to estimate the number of water molecules that are not mixed with the DMSO molecules from the results presented in Fig. 9. The line fitting for the binary and ternary systems of the melting temperature dependence on the DMSO fraction (see Fig. 9) gives the following values: $T_{bin}(X_{DMSO}) = -230.3 \cdot X_{DMSO} + 1.7$ and $T_{ter}(X_{DMSO}) = -302.4 \cdot X_{DMSO} + 0.58$, respectively. From the equation $T_{bin}(X_{DMSO}) = T_{ter}(X'_{DMSO})$ and the condition $(N_{DMSO} + N_w)/N_{lip} = 40.8$, it is possible to calculate that $10 \pm 2.6$ H$_2$O molecules per DPPC molecule are never mixed with DMSO and $X'_{DMSO} = 1.32\, X_{DMSO}$. The obtained value of unmixed water molecules is in agreement with a well-known number of water molecules connected with the polar head group of one DPPC molecule $n_{wb}$, which was determined from DSC measurement and is equal to $10 \div 11$ molecules [5]. The bound water has two important properties: a) it is not frozen at the first exothermic transition as was established in [5], b) it is not mixed with DMSO molecules. This fact supports the idea that DMSO molecules do not penetrate to the region of polar head groups. Moreover, the DSC results give evidence that DMSO molecules do not penetrate to the region of bound water (see Fig. 1). DMSO molecules penetrate only to the region of free water molecules and, thus, strongly influence the intermembrane space and decrease the solvent thickness. This phenomena can be assigned to DMSO-induced dehydration of the intermembrane space.

The membrane repeat distance $d$ is connected with the membrane thickness $d_l$ and the number of solvent molecules per DPPC molecule as

$$d = d_l + \frac{2V_w}{A} n_w + \frac{2V_{DMSO}}{A} n_{DMSO} \quad . \tag{2}$$

$n_w = n_{wb} - n'_w + n_{wf}$ is the number of bound water molecules in the intermembrane space $n_{wb} - n'_w$ and free water molecules $n_{wf}$ (see Fig. 1). Here, $n'_w = 2.2$ is the number of bound water molecules embedded in the region of the polar head groups [7], $n_{wb} = 10 \pm 2.6$, $n_{DMSO}$ is the number of DMSO molecules per DPPC molecule in the intermembrane space, $V_w = 29.9$ Å$^3$ and $V_{DMSO} = 118$ Å$^3$ are the volumes of water and DMSO molecules at $T = 20$°C, respectively. $A = 47.9$ Å$^2$ is the area of one DPPC molecule at the membrane surface in the gel $L_{\beta'}$ -phase [7], and $d_l = 49.6 \pm 1$ Å is a value which is not dependent on the DMSO fraction. $n_{DMSO}$ can be easily calculated from the DMSO mole fraction, assuming that the DMSO mole fraction in the extramembrane space $X'_{DMSO}$ is equal to the DMSO mole fraction in the "free" intermembrane solvent

$$n_{DMSO} = \frac{X'_{DMSO}}{1 - X'_{DMSO}} n_{wf} \quad . \tag{3}$$

From equations (2), (3), and the experimental DSC result, which concerns the constant value of the amount of bound water $n_{wb} = 10 \pm 2.6$, the decrease in $d$ in the region $0.0 \leq X_{DMSO} \leq 0.3$ can be explained as a decrease in the number of free water molecules in the intermembrane space.

The line interpolation of the $d(X_{DMSO})$ function (see Fig. 6) in the region $0.0 \leq X_{DMSO} \leq 0.1$ gives

$$d = 64 - 48.66 \cdot X_{DMSO}. \tag{4}$$

In the region $0.3 \leq X_{DMSO} \leq 0.8$, the repeat distance can be estimated as a constant value of 57 Å. From (4), the value of DMSO mole fraction at which the equilibrium value of $d = 57$ Å can be reached is 0.14. In this case, the decrease in the repeat distance from 64 Å to 57 Å corresponds to a decrease in the number of free water molecules by the value of $2.9 \pm 0.4$.

As a result, the region of small DMSO mole fractions corresponds to the decrease in the number of free water molecules inside the intermembrane space from 4 $X_{DMSO} = 0.0$ to $1.1 \pm 0.4$ at $X_{DMSO} = 0.14$. The DMSO-induced dehydration of the intermembrane space increases the main phase transition temperature from $T_{ch} = 41.7 \pm 0.2$°C at $X_{DMSO} = 0$ to $T_{ch} = 50.6 \pm 0.2$°C at $X_{DMSO} = 0.3$ (see Fig. 6). It is important to note that DMSO and water molecules form hydrogen bonds in the relation 1/2 ($X_{DMSO} = 0.33$) or 1/3 ($X_{DMSO} = 0.25$) [3]. From this approach, the property of ternary system at $X_{DMSO} \geq 0.3$ can be tentatively explained as a result of the strong connection of free water molecules and DMSO molecules in the bulk solvent. At $0.3 \leq X_{DMSO} \leq 0.8$, only bound water exists in the DPPC/DMSO/water system. In this region, the repeat distance $d = 57$ Å and the main phase transition temperature $T_{ch} \approx 52$°C are constant values.

## 6. Conclusions

The properties of dimethyl sulfoxide (DMSO), a cryoprotector that is well known for its biological and therapeutic applications, were investigated for lipid membranes by X-ray diffraction, small-angle neutron scattering, and differential scanning calorimetry.

The more important changes in the property and structure of the ternary DPPC/DMSO/water system occur at small DMSO fractions of $0.0 \leq X_{DMSO} \leq 0.3$. In this region, the main phase transition temperature increases from $41.7 \pm 0.2$°C to $41.7 \pm 0.2$°C, the pre-transition disappears, and the thickness of the intermembrane solvent $d_w$ decreases from $14.4 \pm 1.8$ Å at $X_{DMSO} = 0.0$ to $7.8 \pm 1.8$ Å at $X_{DMSO} = 0.1$. At $0.3 \leq X_{DMSO} \leq 0.8$, the repeat distance $d = 57$ Å and the main phase transition temperature $T_{ch} \approx 52$°C are constant values. In the region of $0.9 \leq X_{DMSO} \leq 1.0$, the DPPC membrane undergoes a transition towards a phase characterized by a small repeat distance of $52 \pm 0.8$ Å Å and a high transition temperature of $74.2 \pm 0.2$°C.

For $0.0 \leq X_{DMSO} \leq 0.3$ the DMSO molecules do not penetrate to the region of the polar head groups of DPPC membrane and, thus, do not influence the membrane thickness. This was established from the SANS experiments. Moreover, DMSO molecules do not penetrate to the region of the bound water as established from the DSC experiment. The decrease in the intermembrane space under the influence of DMSO was determined from complementary SANS and X-ray diffraction experiments.

The DMSO molecules strongly interact with water molecules. This interaction is more intensive than the interaction between the DMSO and the polar head groups of DPPC membrane. The DMSO molecules penetrate neither to the region of the polar head groups nor to the region of bound water, nevertheless, they strongly influence the intermembrane solvent thickness and, as a result, create the DMSO-induced dehydration of the intermembrane space. The number of free water molecules in the intermembrane space was calculated from the common X-ray diffraction, SANS, and DSC results. At room temperature, the number of free water molecules in the intermembrane solvent decreases with increasing DMSO mole fraction and, at $X_{DMSO} = 0.14$, only $1.1 \pm 0.4$ water molecules per DPPC molecule exist without connection to the polar head groups.

## 7. ACKNOWLEDGMENTS

The authors are very grateful to Dr. Sylviane Lesieur (URA 1218 of CNRS) for her help in preparation of the sample. One of us (M. Kiselev) thanks the CEA Saclay for the grant which allowed these measurements to be performed.